\documentclass[aps,
amssymb,
amsmath,
nofootinbib,
superscriptaddress]{revtex4-1}
\usepackage{textcomp}
\usepackage{amsmath}
\usepackage{amssymb}
\usepackage%
{graphicx}
\usepackage{extdash}
\usepackage{epsfig}
\usepackage{textcomp}

\renewcommand{\vec}[1]{\boldsymbol{#1}}
\newcommand{\ensuretext}[1]{\ensuremath{\text{#1}}}

\newcommand{\textmutext}{\ensuretext{\textmu}}
\newcommand{\unit}[1]{\ensuretext{\textrm{\,}}\ensuremath{\mathrm{#1}}}
\newcommand{\eV}{\textit{e}\mathrm{V}}
\newcommand{\MeV}{\mathrm{M}\eV}
\newcommand{\keV}{\mathrm{k}\eV}

\newcommand{\Mum}{\ensuretext{\textmu}\ensuremath{\mathrm{m}}}
\newcommand{\mum}{\textrm{\,\ensuremath{\mathrm{\Mum}}}}

\begin{document}
\title{Using XFELs for Probing of Complex Interaction Dynamics of Ultra-Intense Lasers with Solid Matter}

\author{T. Kluge}
\email[]{t.kluge@hzdr.de}
\affiliation{Helmholtz-Zentrum Dresden-Rossendorf e.V., D-01328 Dresden, Germany}

\author{C.~Gutt}
\affiliation{Deutsches Elektronen-Synchrotron DESY, D-22603 Hamburg, Germany}

\author{L.~G.~Huang}
\affiliation{Helmholtz-Zentrum Dresden-Rossendorf e.V., D-01328 Dresden, Germany}
\affiliation{Shanghai Institute of Optics and Fine Mechanics, Chinese Academy of Sciences, Shanghai 201800, China}
\author{J.~Metzkes}
\affiliation{Helmholtz-Zentrum Dresden-Rossendorf e.V., D-01328 Dresden, Germany}

\author{U.~Schramm}
\affiliation{Helmholtz-Zentrum Dresden-Rossendorf e.V., D-01328 Dresden, Germany}
\affiliation{Technische Universität Dresden, D-01062 Dresden, Germany}

\author{M.~Bussmann}
\affiliation{Helmholtz-Zentrum Dresden-Rossendorf e.V., D-01328 Dresden, Germany}

\author{T.~E.~Cowan}
\affiliation{Helmholtz-Zentrum Dresden-Rossendorf e.V., D-01328 Dresden, Germany}
\affiliation{Technische Universität Dresden, D-01062 Dresden, Germany}
\date{\today}
\begin{abstract}
We demonstrate the potential of X-ray free-electron lasers (XFEL) to advancethe understanding
of complex plasma dynamics by allowing for the first time nanometer and femtosecond resolution
at the same time in plasma diagnostics. Plasma phenomena on such short timescales are of high
relevance for many fields of physics, in particular in the ultra-intense ultra-short laser interaction
with matter. Highly relevant yet only partially understood phenomena may become directly accessible
in experiment. These include relativistic laser absorption at solid targets, creation of energetic
electrons and electron transport in warm dense matter, including the seeding and development of
surface and beam instabilities, ambipolar expansion, shock formation, and dynamics at the surfaces
or at buried layers. \\
We demonstrate the potentials of XFEL plasma probing for high power laser matter interactions
using exemplary the small angle X-ray scattering technique, focusing on general considerations for
XFEL probing.
\end{abstract}
\maketitle
\section{\label{Sec:Introduction}INTRODUCTION}
Free-electron laser sources can provide ultra-short~\cite{Schneidmiller2010,Byrd2010} and intense~\cite{Geloni2010,Amann2012} coherent hard photon beams of up to
several $\unit{\keV}$ (XFEL). Such 4th-generation sources are just currently starting user operation (LCLS, USA~\cite{Emma2010}; SACLA, Japan~\cite{Pile2011}) or will become operational in the near future (Swiss-FEL, Switzerland~\cite{Oppelt2007}; European XFEL, Germany~\cite{Schwarz2004,Schneidmiller2010}). 
Such sources exhibit nearly full transverse coherence in the order of several tens of microns with a high coherence
fraction, eg. $>0.9$ at $10\unit{\keV}$ photon energy, allowing for coherent diffraction and scattering experiments. Focusibility
generally is very good and can reach spot sizes below $1\unit{\mum}$, and possibly less than $100\unit{nm}$ using refractive Beryllium
lenses~\cite{Schropp2013}. When combined with ultra high intensity (UHI) laser systems in the visible or near infrared range, these
sources have the potential of fast and high resolution probing of UHI laser created high density plasmas (HDP) at
the surface and inner regions of a solid. Such plasmas are created when a UHI laser irradiates a solid and quickly
ionizes it due to the extreme laser fields~\cite{Tong2005}, typically reaching field strengths of several TV/m. Current state of the
art measurement techniques – such as proton radiography, phase contrast imaging, reflectometry, Schlieren imaging,
measurement of harmonics, or imaging of the self emission including $K\alpha$ imaging – all suffer from a mismatch of
either their spatial or temporal resolution, or both.

Especially for ultra-high laser intensities, the plasma dynamics becomes highly relativistic and non-linear. 
Short and small scale instabilities are expected to become more important, demanding for novel diagnostics. 
In this paper we will propose techniques that, together with the ultra-short time duration of XFELs, will allow probing of such plasmas on femtosecond time scales otherwise experimentally not accessible. 
For example, nanometer resolution will for the first time enable the direct imaging of high density plasma wave correlations inside the HDP and surface plasmons. 
This may have tremendous impact on the understanding of laser absorption at solid targets, creation of energetic electrons and electron transport in warm dense matter, including the seeding and development of surface and beam instabilities, ambipolar expansion, shock formation, and dynamics at the surfaces or at buried layers. 
The fast probing of such plasmas with XFELs may therefore be important for example in the field of laser fusion~\cite{Tabak1994,Lindl1995,Kodama2001,MosesFusionNIF} or laser particle-acceleration in the radiation pressure (light-sail) acceleration regime (RPA)~\cite{Pegoranov-PhotonBubbles,Henig2009a,Palmer2012,Steinke2013} as well as the target normal sheath acceleration~\cite{Wilks2001,Zeil2010,Gaillard2011,Perego2012,Kim} or shock acceleration~\citep{Forslund1971,Silva2004,DHumieres2005,Fiuza2012}, isochoric heating~\cite{Sentoku2007,Akli2008,LGHuang} and high harmonic generation (HHG)~\cite{HHGSW}.

\begin{figure}
  \centering
  \includegraphics[width=\textwidth]{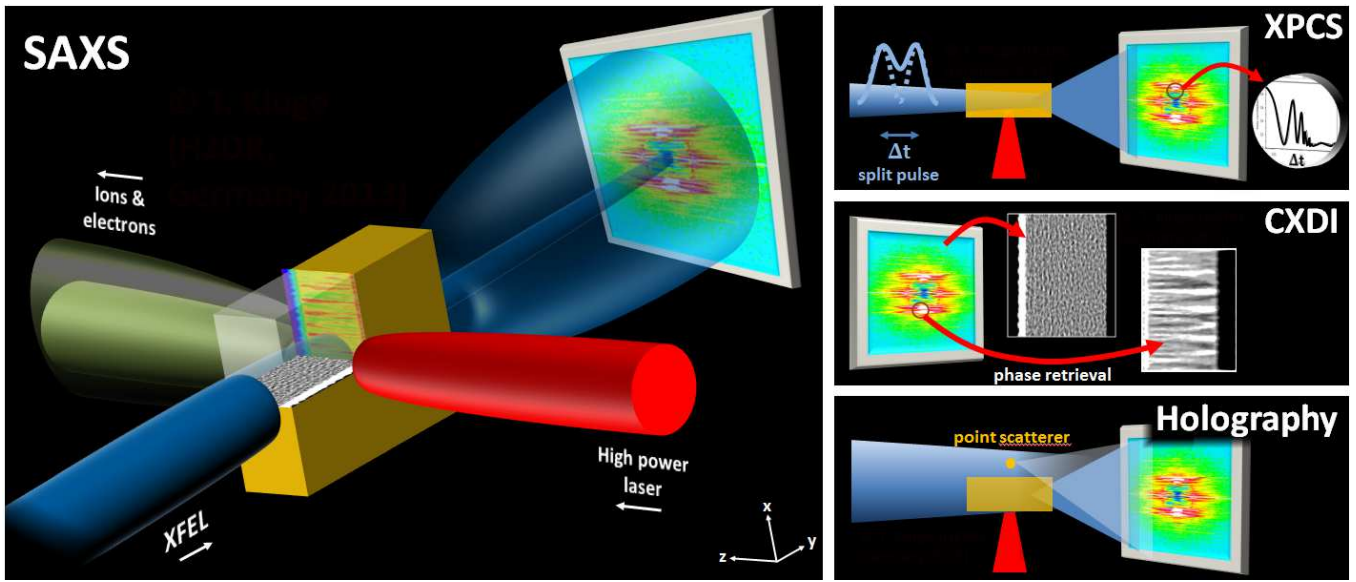}\\
  \caption{SAXS experiment with high power laser. While the UHI laser turns foil into plasma and drives a complex plasma
motion an intense XFEL X-ray beam penetrates the plasma and is scattered to small angles inside the foil and at surface at
high density plasma waves and other density modulations. The SAXS signal shows the density correlations inside the plasma
with few nanometer and few femtosecond resolution. XPCS allows via the analysis of speckle visibility as function of the delay
between two split XFEL pulses to measure the dynamic correlation function. XCDI and holography allow phase retrieval of
the scattered signal. }\label{fig:1}
\end{figure}

The new XFEL light sources provide X-ray beams with photon energy ranging up to $25\unit{\keV}$ (European XFEL). This
enables scattering experiments in a sample with only little absorption. For example in the case of a copper foil, the
absorption length of $8\unit{\keV}$ photons is rather large with $21\unit{\mum}$~\cite{USDepartmentofCommerce}. Spatial modulations of plasma electron density in UHI laser-solid interactions are on the order of $2\pi \mathrm{c}/\omega_p$ ($\omega_p$: electron plasma frequency) which typically is in the range
of few nanometers up to few tens of nanometers, but various other spatial structures with significantly larger scale
lengths may also exist. The scattering angle of such features for $\unit{\keV}$ photons is in the order of a few mrad. Hence,
the UHI laser induced structures generate a density contrast which can be measured by small angle X-ray scattering
technique (SAXS), illustrated in Fig.~\ref{fig:1}. The technique has recently been demonstrated at a seeded XFEL source for
soft matter, but its application to UHI laser produced plasmas would pose new challenges e.g. due to the rapid and complex
dynamics but may open a whole new window for discoveries. In principle the full spatio-temporal information of the
sample electron distribution is contained in the scattered light (holography~\cite{McNulty1992,Lindaas1996,Mancuso2010}) or can be extracted by coherent
X-ray diffraction imaging (CXDI)~\cite{Miao1999,Mancuso2010} by phase reconstruction by iterative phase retrieval ~\cite{Fienup1982}). \\
As another possible diffraction imaging technique that could be developed we propose to use a kind of resonant
coherent X-ray diffraction imaging~\cite{RCXDI}, where the change of the refractive index of the plasma induced by resonant
transitions between two electronic ion states (either for imaging or scattering). The energy at which those resonances
occur does sensitively depend on the chemical element and degree of ionization~\cite{Son2011}. By tuning the XFEL laser
frequency to exactly match this resonant frequency one could use RXCDI to sensitively image the spatio-temporal
distribution of the respective charge state, which – since ionization inside the over-critical plasma is determined by
the electron dynamics – in turn again contains valuable information of the electron dynamics during acceleration and
transport through the plasma.

In addition to scattering experiments various other imaging techniques may be envisioned for use at XFELs for
probing certain properties of UHI laser produced plasmas, e.g. imaging of the polarization change induced by magnetic
fields inside the plasma or at its surfaces (Faraday rotation) or X-ray absorption spectroscopy (XAS), e.g. X-ray
Absorption Near Edge Structure (XANES) or Extended X-Ray Absorption Fine Structure (EXAFS). UHI laser
created electron current filaments going through the target foil give rise to a strong magnetic field around them which
turns the polarization of a probe XFEL beam by an angle in the order of $\unit{\textmutext rad}$ for spatial scales in the order
of a few nanometer expected for solid density plasmas. Employing the coherent nature of XFEL radiation the spatial
distribution of such filaments can be imaged. \\
With XAS on the other hand the spatial distribution of the energy dependent absorption coefficient could be imaged.
The inferred spacing of ions can provide information about the local ion density, chemical environment (geometric
ion structure) and ion dynamics when performed as pump-probe experiment.

In the following we demonstrate the prospects of coherent X-ray diagnostics coupled to UHI laser systems with
special focus to the SAXS method, since this promises to be the easiest to be realized experimentally. We briefly
identify the various types of information that can be obtained and show how this will be of high value as a diagnostic
tool in UHI laser experiments in the future.

\section{\label{Sec:Methods}METHODS}
\begin{table}
  \centering
  \includegraphics[width=14cm]{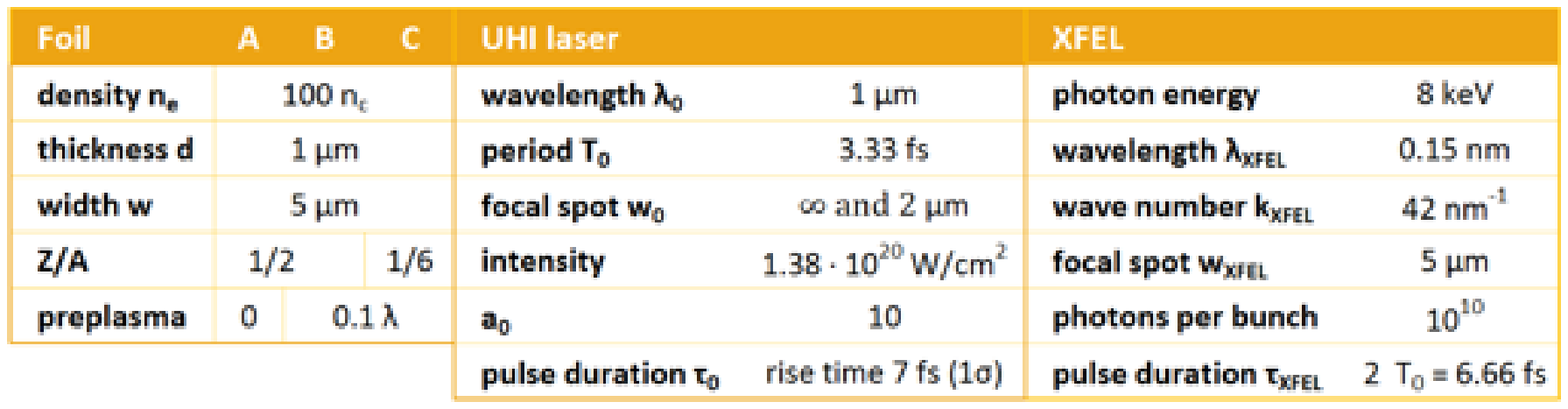}\\
  \caption{Overview of plasma and laser parameters used for the simulations. }\label{tab:parameters}
\end{table}
In this paper we will simulate SAXS experiments at UHI laser-matter interaction experiments and analyze this data
to establish feasibility and value of this new technology. For this purpose we analyze synthesized data of a typical
laser ion-acceleration experiment. The parameters used for the simulations and calculations are summarized in Tab.~\ref{tab:parameters}.

The simulation box size is $10\lambda_0\times10\lambda_0$ with 192 cells per $\lambda_0$ and 192 timesteps per UHI period $T_0$.
The UHI laser is modeled by a spatially plane wave with a short gaussian rise with a 1 $\sigma$-width of $2\lambda_0/\mathrm{c}$, wavelength $\lambda_0$, peak intensity $I_0\unit{[W/cm^2]}=1.38\times 10^{20}/(\lambda_0\unit{[\Mum]})^2$.
This intensity corresponds to a  where $\mathrm{n}_c=\mathrm{m}_e\epsilon_0\omega_0^2e^2$ is the critical electron density for the laser with angular frequency $\omega_0$.  
Typical values for UHI lasers are $\lambda_0=1\unit{\mum}$, $T_0=3.3\unit{fs}$ and $\omega_0=1.9\times10^{15} \unit{Hz}$ resulting in a critical density of $1.1\times 10^{21} \unit{cm}^{-3}$.
Later in the paper we will discuss effects of a finite transverse pulse width.

The flat target foil is modeled by a fully preionized plasma slab of 2λ0 thickness. It consists of ions with charge-to-mass
ratio 1/2 and neutralizing electrons with density $100\unit{n}_c$. The front of the foil is modeled by a step-like density
gradient for run A and was covered by an exponentially decreasing preplasma with scale length $0.1\lambda_0$ for run B. A
third simulation run C was performed with heavier ions of charge-to-mass ratio 1/6. In that run, due to the low ion
mobility, the ion motion can be neglected on the timescale of a few UHI laser periods and the average plasma density
therefore will remain quasi stationary. For all runs the UHI laser is aligned normal to the target foil surface. We
simulate only the plane defined by the laser propagation direction (z-axis) and the electric field polarization (x-axis),
assuming invariance of the system in y-direction.

We compute the scattering signal of the XFEL X-ray beam caused by the fluctuations in the free electron density
assuming an $8\unit{\keV}$ XFEL pulse impinging perpendicular to the simulation plane (x-z-plane) with a beam size of
$5\unit{\mum} × 5\unit{\mum}$ and $10^{10}$ photons per pulse with duration τXFEL (for Fig.~\ref{fig:saxs} and~\ref{fig:saxs} and the corresponding discussion we
adopted $\tau_\mathrm{XFEL} = 2\unit{T}_0$). We assume a detector positioned $1\unit{m}$ behind the target foil. The pixel size of the detector
is taken to be $20~\unit{\mum}$. The electron density modulations observed in the simulations are quite large with up to few
tens of $\mathrm{n}_c$ (see Fig.~\ref{fig:dens}(a,b)). This, in combination with a spatial scale of the modulations in the range of $10\unit{nm}$, yields
a pronounced small angle X-ray scattering signal which allows to image the induced plasma oscillations.

The SAXS signal is related to the Fourier-transformation $\mathrm{n}(\vec{Q})$ of the electron density modulation $\mathrm{n}(\vec{r})$ via
\begin{equation}
\Gamma(\mathrm{Q})=\frac{\Gamma_0}{A}\vec{r}_0^2 \mathrm{n}(\vec{Q})\mathrm{n}^*(\vec{Q}) d\Omega,
\label{eqn:saxs}
\end{equation}
with $\Gamma_0$ the incident X-ray fluence, $A$ the beam size, $\vec{r}_0=2.82 10^{-15}\unit{m}$  the electron scattering length and d$\Omega$ the solid angle of the detector pixel. 
$\vec{Q}$ denotes the wavevector transfer $\vec{Q}=\frac{4 \pi}{\lambda_\mathrm{XFEL}} \sin(\theta/2)\vec{e}_\theta$ and $\theta$ the scattering angle.
For small values of $Q$ this formula reduces to $Q = 2\pi\theta\lambda_\mathrm{XFEL}^{−1}$ and $\vec{Q}$ lies in the plane perpendicular to $k_\mathrm{XFEL}$. In 
Eq.~\eqref{eqn:saxs} we neglected the scattering form factor of individual electrons and atoms, and the structure factor of individual
electron-electron correlations. This is justified since their dimensions are significantly smaller and, consequently, the
associated scattering angles significantly larger than the relevant scales set by the plasma dynamics. Typical scales
encountered in the UHI laser plasma interaction are summarized in Tab.~\ref{tab:scales} together with the corresponding Bragg
scattering angle for the X-ray photon energy of $8~\unit{\keV}$.

\section{\label{Sec:PICresults}PIC RESULTS}
\begin{figure}
  \centering
  \includegraphics[width=14cm]{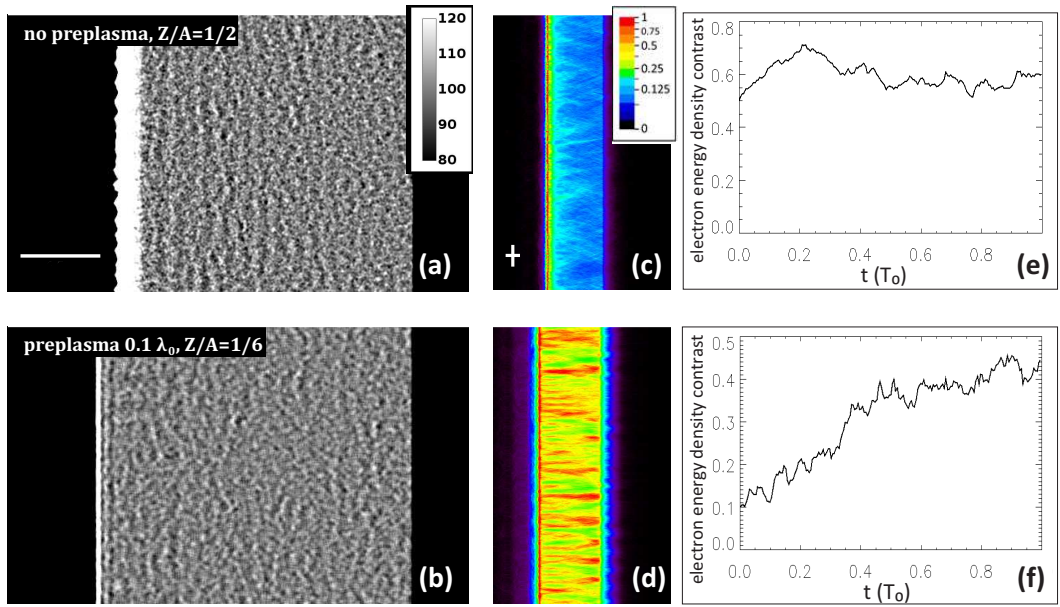}\\
  \caption{PIC results for run A ($L=0$, Z/A=1/2) \textbf{(a,c,e)} and run C ($L=0.1\,\lambda_0$, Z/A=1/6) \textbf{(b,d,f)}, UHI at normal incidence coming from left. \textbf{(a,b)}: Electron density snapshot at $t\approx10\unit{T}_0$ after UHI laser pulse maximum reaches the foil, showing strong surface modulations and high density plasma oscillations, white bar corresponds to $\lambda_0/2$. \textbf{(c,d)}: Electron energy density (pressure) averaged over one UHI cycle starting at same time as (a), (b) and normalized to 2/3 of respective maximum value, white cross corresponds to $\lambda_0/2$. \textbf{(e,f)}: Temporal evolution of hot electron break-up (contrast between maximum and minimum pressure region) when co-moving with an injected bunch. Here, $t=0$ equals the time when the respective electron bunch enters the foil at approx. $10\unit{T}$ after the laser maximum reaches the target. For movie of temporal evolution see supplementary online. }\label{fig:dens} 
\end{figure}

\begin{figure}
  \centering
  \includegraphics[width=10cm]{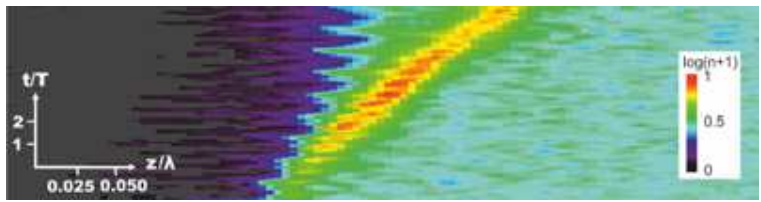}\\
  \caption{Temporal evolution of parametric instability in simulation run A. The normalized electron density along the laser axis as a function of time is found to oscillate weakly with $2\omega_0$ and stronger with $1\omega_0$ along the laser direction. This is expected for parametric decay of the $2\omega_0$ oscillations into two counterpropagating transverse surface waves with frequency $\omega_s=2\omega/2$. }\label{fig:osci}
\end{figure}

We start our analysis with a discussion of the PIC simulation results. Quickly after the onset of the UHI laser
interaction with the foil the laser ponderomotive force pushes the electrons at the front surface into the foil, causing
a steepening of the density profile~\cite{Kemp2008,Chrisman2008,Mishra2009}, see Fig.~\ref{fig:dens}(a). 
When the laser reaches the critical density surface, it is partly reflected within a skin depth of the plasma (moving mirror).
The strong $\vec{v}\times\vec{B}$ fields are able to extract electrons from the skin layer and pull them into the vacuum region~\cite{Gibbon1992,Mulser2008}.
This causes the surface to oscillate along the z-direction with approximately twice the laser frequency. 
This can be seen for example in Fig.~\ref{fig:osci} where the density along the laser propagation direction is plotted as a function of time. 
Some energetic electrons outrun the oscillation and escape into the solid plasma where they continue to move almost ballistic due to the shielding of the laser field.
From the simulations we can infer the approximate temperature for the hot electron component by computing $T^{hot}_{PIC}\cong\left\langle \gamma\right\rangle_{\gamma_0}-\gamma_0$, where $\left\langle \gamma\right\rangle_{\gamma_0}$ is the average energy of electrons with an energy exceeding the threshold energy $\gamma_0$.\footnote{For a single temperature Maxwellian distribution the choice of 
$\gamma$ is arbitrary, but for the situation present at ultra-relativistic laser
intensities 
$\gamma_0$ must be chosen larger than the colder temperatures also present inside the plasma. We choose 
$\gamma_0 = \sqrt{2}$ so that $\mathrm{m}_{e,0}c = p_e$.}
The results for the three simulation runs are $T_{PIC,A}^{hot}=0.41\unit{\MeV}$ (run A) $1.4\unit{\MeV}$ (run B), $1.3\unit{\MeV}$ (run C). 
This compares very well with analytical models~\cite{Kluge2011a}
\begin{equation}
T^{hot}\cong \mathrm{m}_e\mathrm{c}^2 \left(\frac {\pi}{2K\left(-\hat{a}^2\right)}-1\right)
\label{eqn:thot}
\end{equation}
where $K$ is the complete elliptical integral of the first kind, and $\hat{a}$ is the normalized laser field strength amplitude at
the critical density surface. Inferring $\hat{a}$ from a (non-relativistic) calculation considering an exponential preplasma~\citep{Rodel2012} we obtain for the parameters\footnote{The effective scale lengths differ from the initial setup and is for run A approximately $5\unit{nm}$ while for run B, C the density at the surface follows more a two-exponential decay around the relativistic critical density, with approximate scale lengths $L_{short}\cong 15\unit{nm}$ and $L_{long}\cong 70\unit{nm}$. } of run A ($\hat{a}\approx 3$) $T^{hot}_A\cong 0.42\unit{\MeV}$. 
For runs B, C we obtain depending on the exact effective scale length $T^{hot,short}_{B,C}\cong 1.3\unit{\MeV}$ with $\hat{a}\approx 8$ (for $L=L_{short}$) and $T^{hot,long}_{B,C}\cong 1.8\unit{\MeV}$ with $\hat{a}\approx 11$ (for $L=L_{long}$) in good agreement with the simulations. 

A closer look at the electron dynamics reveals a rather complex situation. 
For example, a non-uniform spatial distribution of the hot electrons can be observed (Fig.~\ref{fig:dens}(c,d)), namely due to surface structures and filamentation inside the target, despite the completely homogeneous 1D initial setup. 
Distinct differences between run A and C can be seen with respect to the nature of those electron break-up processes. 
While in run A the hot electrons show a transverse structure directly at the foil surface, in run C the filamentation sets in only during the first few tenths of a laser wavelength $\lambda_0$ (Fig.~\ref{fig:dens}(e,f)). 
In run A the filamentary structure is due to a parametric two plasmon decay
instability ~\cite{Sentoku2000,Macchi2001,Macchi2002}.
Here, additionally to the $2\omega_0$ oscillation of the critical surface one can observe a $1\omega_0$ oscillation
with a transversely modulated amplitude (Fig.~\ref{fig:osci}). This slow oscillation is a direct consequence of the superposition
of two surface plasmons originating from the decay of the $2\omega_0$ oscillation. More details of this two-plasmon decay at
high plasma density will be published elsewehere. \\
In run C the electron filamentation is due to a Weibel-like instability and the filaments are accompanied by strong
quasi-static magnetic fields perpendicular around them. While in run C the foil front surface remains flat throughout
the simulation due to the low ion charge-to-mass ratio and therefore low ion mobility, in run B (not shown) the front
surface evolves similar to run A and also becomes structured. Yet, the filamentary structure of the energetic electron
inside the foil is distinctly different to run A – and very much the same as in C – with respect to spatial scale lengths,
modulation strength, and directionality. For run B and C the filaments are aligned almost parallel to each other and
perpendicular to the front surface. The average distance between filaments is in the order of $0.25\lambda_0$ at the target rear.
In this sense, run B represents an intermediate case between run A (dominated by parametric instability) and run C
(dominated by Weibel-like instability). 
In experiment the specific properties of the filaments are determined by the
details of the initial setup and laser properties. Additionally a number of other types of instabilities and filamentation
mechanisms may occur, each with distinct properties. For example, these include resistive~\cite{Sentokuresistive}, electro-thermal~\cite{HainesInstability},
and Rayleigh-Taylor-like~\cite{Palmer2012} instabilities. Probing these instabilities is crucial to optimize the experimental conditions
for applications, as they can significantly influence the laser absorption and plasma evolution. \\
To illustrate this point, we have performed additional simulations with preformed structure on the foil surface, finding
that e.g. the spacing of filaments is sensitive to frequency and amplitude of these structures. Yet, presently, due to
the impossibility of diagnosing them in-situ inside the solid density plasma with sufficiently high spatial and temporal
resolution (sub- μm and few fs), experiments have always relied on secondary processes like the imprint of the electron
density distribution on subsequently accelerated ions~\cite{Palmer2012,Fuchs2003,Metzkes}, and Kα imaging is limited to long-scale (spatially
and temporally) features of thick targets~\cite{Gaillard2011}. Additionally, we see in the simulations that during the electron passage
through the foil the electron channels tend to merge. Consequently, sampling only the density distributions at the
surfaces, as in the case when detecting ion, is not sufficient for identifying the nature of the structures. Such indirect
techniques neither provide sufficient temporal nor spatial resolution along the laser propagation axis and one has
to rely on propagation models or simulations to infer the electron distribution inside the target from the measurements. 

Generally speaking, understanding filamentation is crucial for an understanding of the laser energy absorption
processes, since they can significantly increase or inhibit energy absorption and therefore have a tremendous impact
on the dynamics at the front surface (rippling, hole-boring and compression), electron transport inside the target and
its imprint to ion acceleration (e.g. filamentation of TNSA ion beam~\cite{Metzkes}, instable RPA~\cite{Pegoranov-PhotonBubbles}) and HHG~\cite{HHGSW}. Furthermore their
presence is almost unavoidable in relativistic laser-solid interaction. This physics is therefore one important example
out of many where XFEL X-ray scattering may prove helpful. For this reason we expand in the following on this
example in order to demonstrate the power of the XFEL probing. We show that with X-ray scattering at XFELs
temporally and spatially resolved measurements of these processes come into reach.

\section{\label{Sec:SAXSresults}SAXS RESULTS}
\begin{table}
  \centering
  \includegraphics[width=10cm]{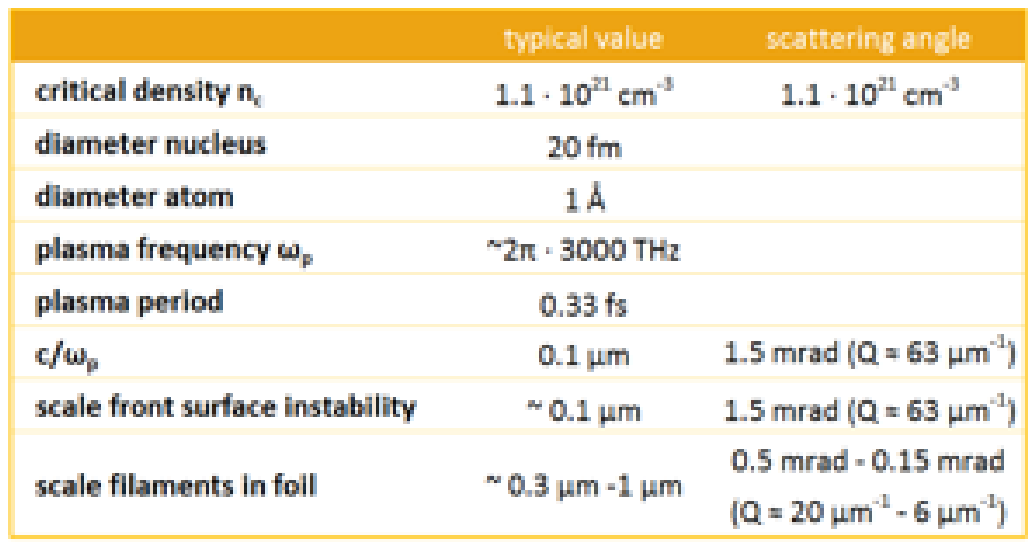}\\
  \caption{Typical temporal and spacial scales for simulations and parameters used in this paper together with corresponding Bragg scattering angles $\varphi = \mathrm{asin}\frac{\lambda_{XFEL}}{S}$ ($S$: scale length) for $8\unit{\keV}$ photons. }\label{tab:scales}
\end{table}
Based on the PIC results we compute the respective expected SAXS signal on a detector 1m behind the target foil
using Eq.~\eqref{eqn:saxs}. The plasma electron density modulations are strong enough to to yield even for the scattering from
the interior of the foil a number of scattered X-ray photons per pixel on the detector of well above 1. The resulting
SAXS pattern $\gamma(\vec{Q})$ at a fixed point in time is shown in the upper insets in Fig. 4.
However, these figures are SAXS images from a density snapshot only. This would correspond to an XFEL pulse
duration much less than the smallest dynamic timescale of the plasma evolution, $\tau_\mathrm{XFEL}\ll \omega_p^{−1}$. Yet, for realistic
XFEL pulses the pulse duration $\tau_\mathrm{XFEL}$ will be in the order of a few femtoseconds to a few tens of femtoseconds while
$2\pi/omega_p$ for solid foils is in the order of a few hundred attoseconds. In such a scenario the resulting SAXS signal $\tilde{\Gamma}(\vec{Q})$ is composed of a superposition of the individual scattering signals $\Gamma(\vec{Q},t)$,
\begin{equation}
\tilde{\Gamma}(\vec{Q})=\int_t^{t+\tau_{XFEL}} I(\vec{Q},t) dt.
\end{equation}
Here, $\tilde{\Gamma}(\vec{Q})$ is the X-ray fluence observed by a detector pixel during an XFEL pulse of duration $\tau_{XFEL}$. 
Fig.~\ref{fig:saxs} depicts
the respective calculated SAXS images computed from adding the individual Fourier transforms of each timestep of
the PIC simulations over two UHI laser oscillations (for $\lambda_0=1\unit{\mum}$ this corresponds to approximately $\tau_\mathrm{XFEL} = 6.7\unit{fs}$).
As one would expect, the main difference compared to the SAXS images from a single timestep is the reduced speckle
contrast while the main scattering features described below remain visible. The reason is the long correlation time of
the UHI laser induced density fluctuations compared to the XFEL pulse duration, even though the timescale of those
fluctuations itself is significantly shorter in the order of the plasma period $\omega_p=\left(n_e e^2/m_e\varepsilon_0\right)^{1/2}$. \\
The smearing of the speckle pattern due to the comparably fast plasma dynamics makes a real space phase retrieval
approach for direct imaging difficult with XFEL pulses much longer than $\unit{T}_0$. Yet, this effect can be used to track the
dynamics of the plasma modulations by X-ray photon correlation spectroscopy (XPCS). Details are described later
in section~\ref{SubSec:Dynamics}. Other important quantities deducible from the scattered intensity images are the form factor $F(\vec{Q})$ and structure factor $S(\vec{Q})$.

\subsection{\label{SubSec:Form}Form and structure factor}
\begin{figure}
  \centering
  \includegraphics[width=\textwidth]{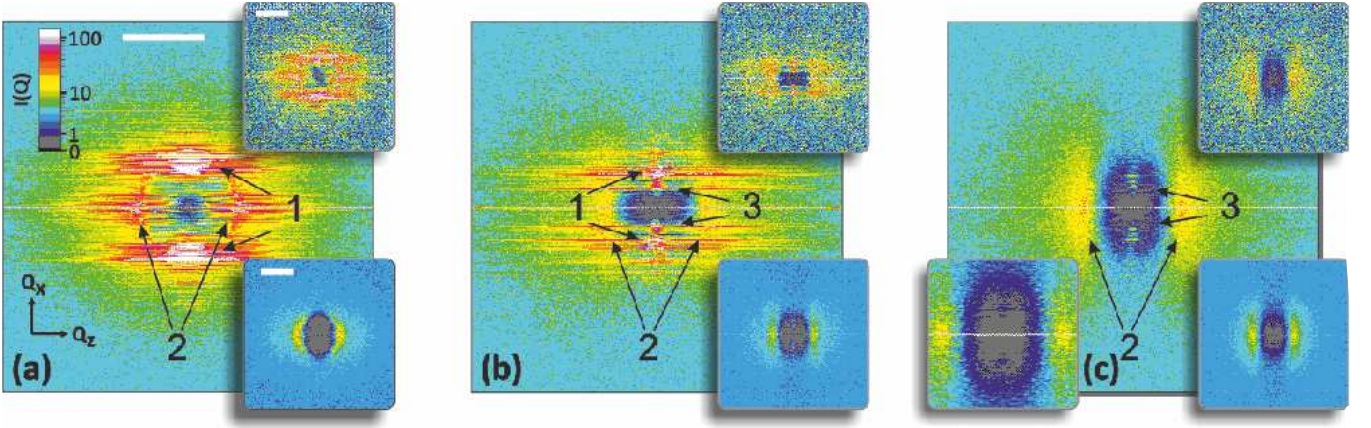}\\
  \caption{Simulated SAXS images based on the PIC results for run A-C (a-c). The main figures show the superposition of $\tilde{\Gamma}(\vec{Q})\cdot A_\mathrm{pixel}$ ($A_\mathrm{pixel} = 20 \times 20\unit{\mum}^2$ is the detector pixel area) calculated from 384 individual timesteps (2 UHI periods) starting approx. $9\unit{T}_0$ after the UHI laser pulse maximum has reached the foil. Lower insets show the corresponding signal excluding the front surface (lower left inset in c is a magnification of central SAXS region). Upper insets show the signal from a single timestep, corresponding to an infinitely short XFEL pulse. The origin $\vec{Q}=0$ lies in the center of each picture, white bars correspond to $0.1\unit{nm}^{-1}(\lambda_0\,[\Mum])^{-1}$. Color scale is logarithmic, representing estimated number of X-ray photons per detector pixel assuming the XFEL pulse defined in Sec.~\ref{Sec:Methods}, $\lambda_0=1\unit{\mum}$, sample depth $5\unit{\mum}$ in $\vec{Q}_x$-direction, detector pixel solid angle corresponding to $20\unit{\mum}\times 20{\mum}$ at a distance from sample of $1\unit{m}$. }\label{fig:saxs}
\end{figure}

Before discussing how in certain cases the form factor and structure factor can be determined individually we will
first discuss the product of the two, $\tilde{\Gamma}(\vec{Q})$, Fig.~\ref{fig:saxs}.\\
The most obvious signals are the streaks in the direction parallel to the surface normal representing the jump in
the electron density at the surface of the sample. A sharp jump in density gives broad lines in the Fourier transform.
Consequently, these lines are absent when considering only the signal produced by the interior of the foil (see lower
insets). In contrast to the un-driven surface, the driven system of run A shows a high degree of surface roughness
with strong modulations along the vertical direction. This leads to a strong reduction in the X-ray reflectivity from
the surface along the horizontal direction and to intense lobes (marked "`1"'’ in Fig.\ref{fig:saxs}) of scattering intensity in the
x-direction which allows to track the spatial modulations in the vicinity of the surface. \\
The second important signal are the lobes in the Qz-direction which are the signal from scattering at the high density
plasma oscillations inside the target (marked  "`2"'). As can be seen from Fig.~\ref{fig:dens}, the plasma oscillations make up a
grating-like density pattern inside the foil, aligned parallel to the front surface and oscillating with roughly the plasma
frequency $\omega_p$ which for the given simulation parameters is $\omega_p=10\omega_0$. Despite the short oscillation period compared to the XFEL pulse duration these oscillations lead to a distinct scattering signal due to the long correlation time.
The X-ray photons are scattered by the density modulations in the z-direction and appear in the SAXS signal at a
distance from the origin of roughly $Q_z\approx0.065\unit{nm}^{-1}(\lambda_0\,[\Mum])^{-1}$. For $8\unit{\keV}$ X-ray photons and $\lambda_0=1\unit{\mum}$ this would
correspond to a scattering angle of$1.6\unit{mrad}$. This value coincides with the scattering angle of $1.55\unit{mrad}$
calculated for the XFEL photons being scattered at the cold plasma wave, assuming $k_p\cong c/\omega_p$: $\varphi=\mathrm{asin}\left(k_p/k_{XFEL}\right)$.
Since the plasma is cold initially, and therefore the phase velocity equals zero, it is reasonable to assume that the
plasma waves inside the foil are excited in the wake of the energetic electrons accelerated by the UHI laser at the
front surface and travelling through the foil with $v_z\lesssim c$.

The scattering signal from plasma oscillations therefore contains important information about the laser interaction
with the target and the transport of energetic electrons through it. An example of this fact is the break-up of energetic
electron beams travelling through the foil (filamentation) (Fig.~\ref{fig:dens}(c-f)). The electron oscillations together with the
background density contain this information which leads –- due to the periodic structure of the filaments along the
x-axis -– to scattering in x-direction.
For clarity we first analyze the SAXS signal from the foil interior region only (lower insets Fig.~\ref{fig:saxs}), since here the
scattering from the surface structure is not included. Such a signal could be obtained for an extremely tightly
focused XFEL beam (cp.~\cite{Schropp2013}), sparing the foil front surface region near the UHI laser focus. While in run A there
is no significant filamentation present (Fig.~\ref{fig:dens}(c)), and therefore no pronounced vertical scattering signal is visible,
for run B and C one can clearly see two scattering features along the vertical x-axis. When analyzing the whole
foil sample B, these features remain clearly visible ("`3"') while the signal from diffraction at the front surface ripples
is superimposed ("`1"'). For run C there is no significant difference between the distribution of the scattered X-ray
from the inner foil and scattering from the whole foil since the front surface there remains nearly flat and does
not lead to such additional scattering features. The density preplasma gradient in this case is persistent
due to the inhibition of plasma steepening by the only less mobile ions, which removes the distinct broad lines in
$\tilde{\Gamma}(\vec{Q})$ seen in run A (and less prominently also in run B). From the filamentary scattering signal "`3"' it is possible
to deduce the typical spatial vertical spacing of energetic electron currents. The lowest spatial frequency in vertical
direction is found to be approximately $0.0245\unit{nm}^{-1}(\lambda_0\,[\Mum])^{-1}$ corresponding to a spatial distance of $0.25\lambda_0$, in good
agreement with the results from analyzing the PIC simulated electron energy density patterns (Fig.~\ref{fig:dens}(d)). Referring
to the aforementioned additional simulation runs including a front foil surface roughness, we would like to note that we
checked that the lowest spatial frequencies in vertical direction in the simulated SAXS signal consistently correspond
to the approximate vertical spacing of hot electron filaments. It hence can be concluded and it can be shown\footnote{This can be seen by comparing the absolute values of scattered intensity in the respective $\vec{Q}$ region for the sum of the Fourier transforms
from multiple timesteps with the Fourier transform of the time averaged density.} that
the energetic electron break-up leads to modulations of the average density as well as phase interfaces of individual
plasma oscillations excited by individual electron filaments, which both give rise to a corresponding signal in the (time
integrated) Fourier transform $n(\vec{Q})$.

\begin{figure}
  \centering
  \includegraphics[width=\textwidth]{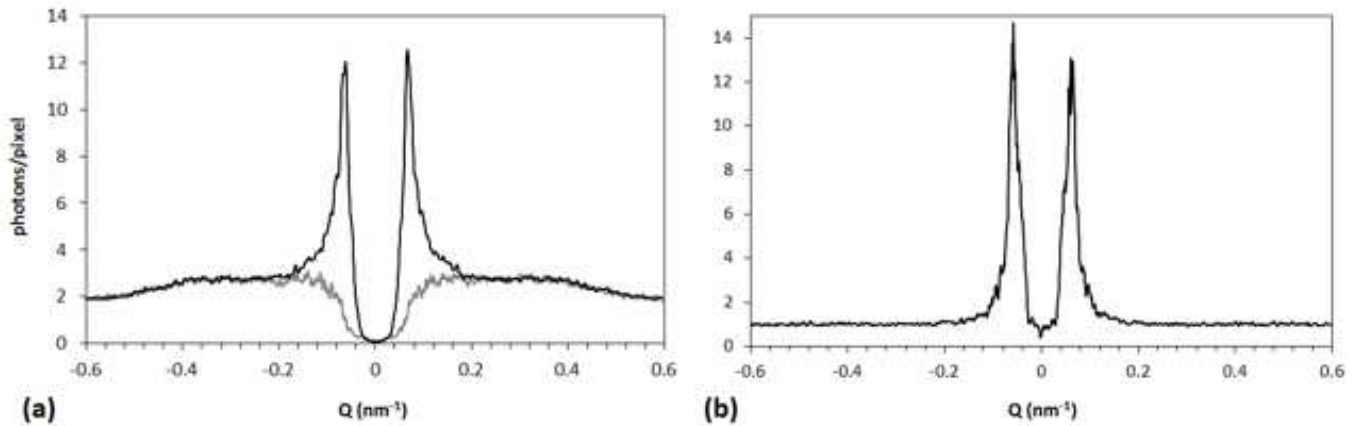}\\
  \caption{(a) Cut through $\tilde{\Gamma}(\vec{Q})$ of run C (Fig.~\ref{fig:saxs}(c)) along the horizontal (black) and vertical (gray) direction next to the center (in $Q_x$-direction the cut is aligned far enough to omit the vertical scattering features). Cut in vertical direction is dominated by $F(\vec{Q})$, while horizontal cut is composed of the product $F(\vec{Q})S(\vec{Q})$. (b) shows the approximate structure factor in $Q_z$-direction obtained by $S(Q_z)\cong \tilde{\Gamma}(Q_z)/F(Q_x)$.}\label{fig:FS}
\end{figure}
The spatial correlation between the plasma modulations is called the structure factor $S(\vec{Q})$ which leads to the
asymmetric scattering pattern seen in Fig.~\ref{fig:saxs} at small $\vec{Q}$ values. This asymmetry reflects the different degree of spatial
correlation in the plasma: along the laser beam direction, the ordering is much higher than perpendicular to it. The
line width and height of $S(\vec{Q})$ allow to describe the degree of spatial correlations. The form factor $F(\vec{Q})$ describes
the averaged geometrical shape and size of a single density modulation. The overall scattering intensity is a product
of structure and form factor $\Gamma(\vec{Q})=F(\vec{Q}) \times S(\vec{Q})$. Hence, the above discussion of the SAXS images Fig.~\ref{fig:saxs} is based
on the assumption that the scattering features are primarily due to the structure factor $S\vec{Q}$. This is true since,
as will be shown now at the example of the plasma oscillations, the form factor typically exhibits only long period
oscillations in the frequency domain.

In the case of scattering off the horizontal plasma oscillations only, which is realized best in run C (besides the
comparatively weak vertical scattering signal from the filaments), the scattering in the vertical x-direction is predominantly
due to the form factor, $\Gamma(Q_x)\cong F(Q_x)$) since the structure factor is close to unity. In the horizontal direction
both $F(\vec{Q})$ and $S(\vec{Q})$ contribute to the scattering intensity. Horizontal and vertical cuts through the scattering intensity
are shown in Fig.~\ref{fig:FS}(a,b) and demonstrate this asymmetry. The scattering intensity shows long period oscillations
in both horizontal and vertical direction representing the form factor $F(\vec{Q})$ of the plasma oscillations. As the structure
factor in the vertical direction is basically unity, the ratio of horizontal to vertical scattering intensity yields $S(Q_z)$,
as shown in Fig.~\ref{fig:saxs}(c). The high degree of correlation in the plasma waves leads to a strong and rather narrow first
maximum of the structure factor. As assumed above, $S(\vec{Q_z})$ does indeed not qualitatively differ from $\Gamma(\vec{Q_z})$).

\subsection{\label{SubSec:Dynamics}Dynamics}
X-ray photon correlation spectroscopy (XPCS) allows to track the ultrafast dynamics of the plasma modulations
even below timescales of the XFEL pulse duration τXFEL~\cite{Grubel2007}. XPCS techniques at pulsed light sources rely on some
sort of speckle visibility spectroscopy. The idea is to split a single pulse into two by means of a split-and delay X-ray
beam line which allows to adjust the time delay between the two pulses. The two pulses are producing overlapping
speckle patterns on the CCD detector, which is too slow to read them out separately on ultrafast time scales. By
analyzing the speckle visibility as a function of delay time between the pulses, the usual intensity-autocorrelation
function $g2(q,\tau)$ can be retrieved~\cite{Gutt2008}. \\
\begin{figure}
  \centering
  \includegraphics[width=\textwidth]{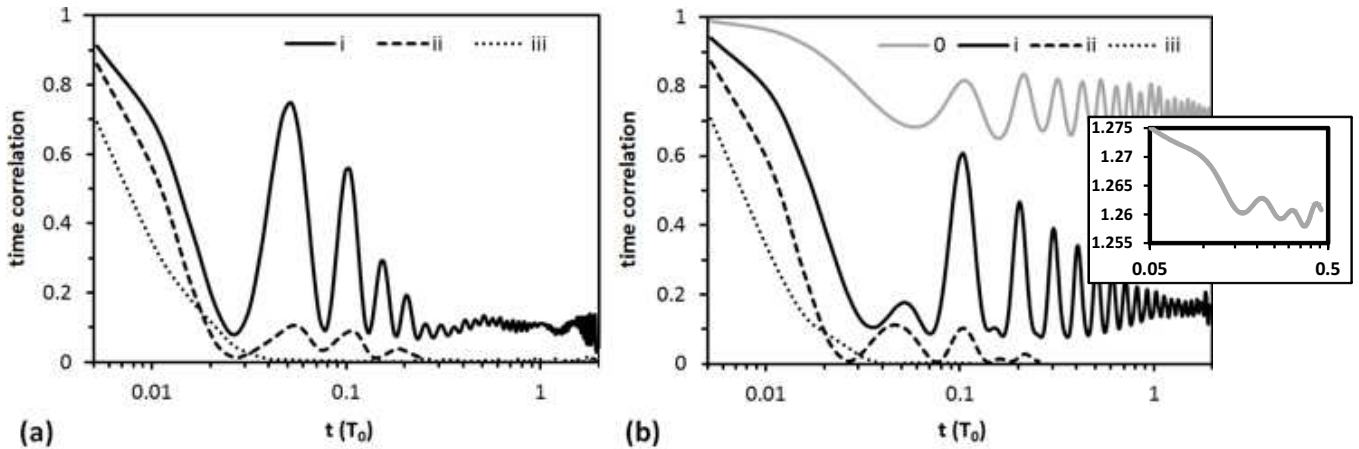}\\
  \caption{Time correlation function $g2(\vec{Q},\tau)$ for simulation run C and $Q=0.035\unit{nm}^{-1}$ (labled "`0"'), $0.066\unit{nm}^{-1}$ (i), $0.144\unit{nm}^{-1}$ (ii), $0.357\unit{nm}^{-1}$ (iii) in the horizontal z-direction (a) and vertical x-direction (b). Feature "`0"' in vertical direction corresponds to the signal from the filaments (”‘3”’), which exhibits a long correlation time. (i) corresponds to $omega_p/c$, (ii) and (iii) are dominated by overdamped waves. The inset shows an exemplary speckle visibility as a function of pulse delay, simulated for
an ultra-short XFEL pulse duration of $\tau_\mathrm{XFEL}\cong0.15\unit{T}_0$ (approx. $0.5\unit{fs}$ for $\lambda_0 = 1\unit{\mum}$). Ranges in $Q_x$ and $Q_z$ of $\pm0.017\unit{nm}^{-1}$ were used for averaging, and a time window of $2~\unit{T}_0$ starting at approx. $10\unit{T}_0$ after the UHI laser pulse maximum reached the foil was considered.}\label{fig:corr}
\end{figure}
The simulations allow to produce such speckle visibility functions by summing up scattering intensities from different
frames. Figure~\ref{fig:corr} shows the $g_2$ functions for different values of the wave vector transfer. At small $Q$ values the correlation
functions are oscillating in time. A temporal Fourier transform of the correlation functions yield the dynamic structure
factor $S(Q,\omega)$ (Fig.~\ref{fig:corr_temp}). By analyzing the oscillation frequency as a function of $Q$ it becomes possible to directly acquire
the dispersion relation of the plasma and the cross-over from inelastic to quasielastic excitation can be monitored.\\
\begin{figure}
  \centering
  \includegraphics[width=8cm]{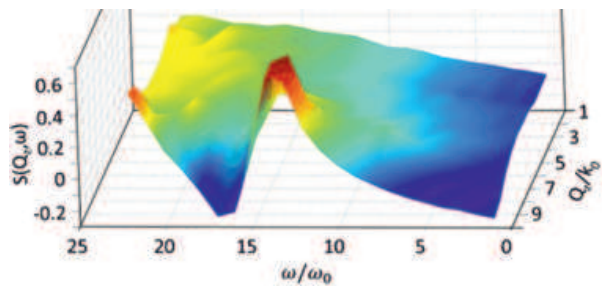}\\
  \caption{Dynamic structure factor obtained by Fourier transformation of $g_2$. For small values of $Q_z$ the plasma oscillations produce a spike in the structure factor while for larger $Q_z$ the cross-over to quasi-elastic signal produced by overdamped waves is observed.}\label{fig:corr_temp}
\end{figure}
As can be seen in Fig.~\ref{fig:corr}(a), along the $Q_x$-axis, corresponding to vertical structure patterns in the electron density
due to electron filaments and front side structure, the oscillation frequency is the plasma frequency, which represents
just the fastest possible plasma dynamics time scale. In Fig.~\ref{fig:corr}(a), for small $Q_z$-values around $Q_x = 0$, the correlation
functions are oscillating with twice the typical frequency of the plasma waves in the sample. This is expected since this
region is governed by scattering at the plasma oscillations. After half a plasma period the sample density modulations
are to a large extent simply inverted. The Fourier transform of the inverted density pattern is the same as the Fourier
transform of the original pattern. Hence the correlation function has a local maximum at $t=2\pi \omega_p^{-1}/2$.
With increasing Q values the oscillations are damping out until at the largest $Q$ values the correlation functions show
an exponential decay only.
It is important to emphasize that the achievable temporal resolution is not strictly limited by the XFEL pulse duration
which in the case of European XFEL operating with self seeding for example will be longer than approximately 2 fs~\cite{Tschentscher2011}.
Rather, it is determined by the time separation of the split beams while the pulse duration determines the average
speckle visibility and therefore the offset and noise level in the temporal speckle visibility function. We show the
speckle visibility function for one example in the inset of Fig.~\ref{fig:corr} and an ultra-short XFEL pulse duration of $0.15\unit{T}_0$
(approx. $0.5\unit{fs}$ for $\lambda_0 = 1\unit{\mum}$). For much longer pulse durations the amplitude of the fast (plasma period) oscillations
of speckle variance would become less than the jitter of the average needed for normalization. Yet, with the future
development, such short XFEL pulses may come into reach and enable XPCS even on the fast timescales of the plasma
oscillation.

\section{\label{Sec5}Discussion and Outlook}
In this section we want to discuss general issues of relevance for a real experiment and show how the above idealized estimates and discussions can be transferred to reality. 

The possibility to use XFEL light as a probe tool for UHI laser experiments in general requires that the scattered
X-ray signal must be strong enough to stand out from any X-ray background present from the plasma self emission.
Otherwise the desired signal might not be differentiated from this data. The self emission from a laser generated
hot plasma, which is originating primarily from electronic transitions to the K-shell, can be quite intense for high
power lasers. The emission duration is a function of the time of hot electron presence and its intensity is linked to the
density of ions and hot electrons. Its absolute quantity therefore depends on parameters such as laser pulse duration,
intensity, focus size but also target thickness and composition~\cite{Neumayer2010}. In this reference, for the
specific parameters used there, the fraction of total energy of the X-rays from self emission to the energy contained
in the laser pulse was found experimentally to be in the order of $10^{-4}$. To estimate the number of photons from self
emission we may expect at the detector in our case, we adopt this number and scale it to $15\unit{J}$ which is a typical value
for short pulse UHI lasers. Assuming $8\unit{\keV}$ for the average photon energy ($K\alpha$ from copper) and isotropic emission,
we expect roughly $12\times10^{11}$ $K\alpha$ photons emitted in the full solid angle and roughly $10$ photons per pixel at the detector position $1\unit{m}$ behind the target with pixel size $20^2\unit{\mum}^2$ as planned for HIBEF at the European XFEL~\cite{HIBEFConsortium}.
Hence we find that the SAXS signal, especial regarding the most intense features, will be well visible with a signal
to noise ratio of up to 15. However, the less intense scattering features are found to be only as intense as the $K\alpha$
noise. Therefore one has to remove the self emission from the signal. Yet, one should be able to almost completely
circumvent this problem by using either Bragg mirrors (HOPG) that by spatial energy dispersion allow filtering of
the narrow $K\alpha$ photons, or by adding an energy selective absorption filter in front of the detector (e.g. a foil made of the same material as the target and tuning FXEL energy slightly below $K\alpha$ energy).

\begin{figure}
  \centering
  \includegraphics[width=\textwidth]{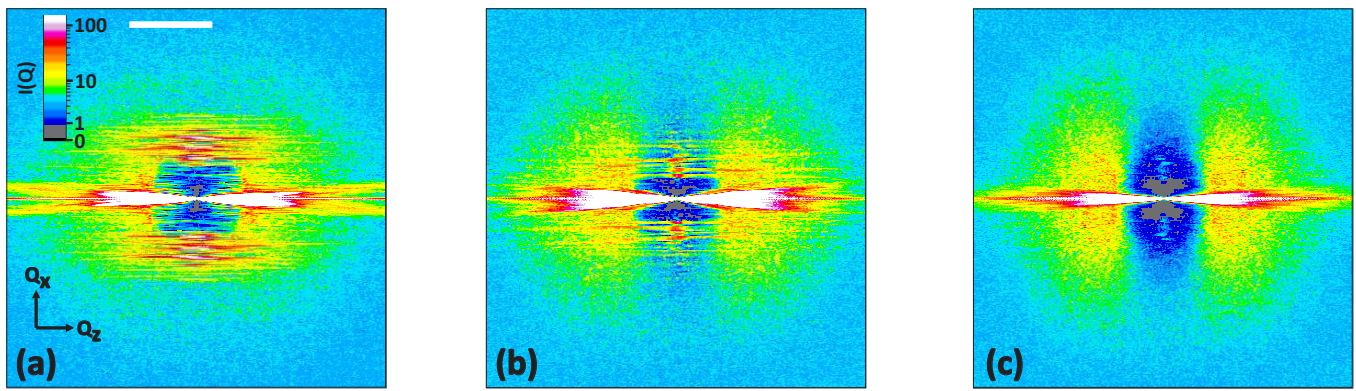}\\
  \caption{Simulated SAXS images based on the PIC results for run A-C (a-c) but with tight UHI laser focus of $2\lambda$. The images
are similar to those with plane UHI laser geometry Fig. 4 with the exception of the dominant signal around the $Q_z$ axis from
the front surface curvature introduced by ponderomotive holeboring. Vertical features from filaments and front surface ripples
remain clearly distinguishable.}\label{fig:saxs_pulse}
\end{figure}

In a real experiment the finite width of the UHI laser pulse and 3D effects have to be considered. Especially imaging
techniques that do not record correlations but rather integrate along the propagation direction, such as for example
Faraday rotation imaging, XAS and others only have a 2D resolution in the plane perpendicular to the XFEL beam.
Correlation techniques on the other side may suffer from a distortion from curved features from the spatial, non-flat
pulse form. Fig. 8 shows the calculated SAXS signal for parameters as before but with a finite UHI laser pulse width
$w_0 = 2\unit{\mum}$ rather than plane wave. The important features discussed before are still prominent and intense while
being overlaid by an intense signal around the $Q_z$ axis from the curved foil surface originating from the curved laser
intensity profile holeboring. We conclude that even in such a case of extremely tight focusing the relevant information
will still be detectable. Going to a 3D geometry one would expect similar effects combined with a moderate reduction
in intensity of the scattered X-ray light due to the reduced thickness of filaments and plasma wave correlation in
y−direction compared to the assumptions made above.
In principle one can work around the 2D and 3D effects in several ways. For example, in the case of XAS or RCXDI, due to
its high chemical sensitivity, sandwich targets – containing only a thin sheet of the respective material of interest aligned perpendicular to the XFEL beam – limit the longitudinal resolution to the thickness of the sheet. If
the surrounding material is chosen such that the electron density and resistivity are similar as inside the sheet, the
compound target will behave similar with respect to the UHI laser interaction and electron transport inside the target.
Another solution can be computed tomography from multi XFEL-beams probing the target from several angles at
once, using a non-colinear split and delay beam line.
We have shown that an XFEL combined with UHI laser experiments is very promising versatile and unique tool
for ultra-fast ultra-small scale resolution plasma probing important for many physical phenomena, such as relativistic
laser absorption at solid targets, creation of energetic electrons and electron transport in warm dense matter, including
the seeding and development of surface and beam instabilities, ambipolar expansion, shock formation, and dynamics
at the surfaces or at buried layers.

\begin{acknowledgments}
This work was partially supported by the joint research project onCOOPtics under Grant No. 03ZIK445.
\end{acknowledgments}

\bibliographystyle{unsrt}
\bibliography{main2}

\begin{thebibliography}{10}

\bibitem{Schneidmiller2010}
{E.A. Schneidmiller} and {M.V. Yurkov}.
\newblock {EXPECTED PROPERTIES OF THE RADIATION FROM THE EUROPEAN XFEL
  OPERATING AT THE ENERGY OF 14 GEV}.
\newblock In {\em Proceedings of FEL2010}, page 119, Malm\"{o}, Sweden, 2010.

\bibitem{Byrd2010}
{J. M. Byrd}, {G. Huang}, and {R. Wilcox}.
\newblock {Femtosecond operation of the LCLS for user experiments}.
\newblock In {\em Proceedings of IPAC’10, Kyoto, Japan}, page 2287, Kyoto,
  Japan, 2010. IPAC'10 OC/ACFA.

\bibitem{Geloni2010}
Gianluca Geloni, Vitali Kocharyan, and Evgeni Saldin.
\newblock {Scheme for generation of fully-coherent, TW power level hard X-ray
  pulses from baseline undulators at the European X-ray FEL}.
\newblock {\em arXiv preprint}, 1007.2743:23, July 2010.

\bibitem{Amann2012}
J.~Amann, W.~Berg, V.~Blank, F.-J. Decker, Y.~Ding, P.~Emma, Y.~Feng,
  J.~Frisch, D.~Fritz, J.~Hastings, Z.~Huang, J.~Krzywinski, R.~Lindberg,
  H.~Loos, A.~Lutman, H.-D. Nuhn, D.~Ratner, J.~Rzepiela, D.~Shu, Yu. Shvyd'ko,
  S.~Spampinati, S.~Stoupin, S.~Terentyev, E.~Trakhtenberg, D.~Walz, J.~Welch,
  J.~Wu, A.~Zholents, and D.~Zhu.
\newblock {Demonstration of self-seeding in a hard-X-ray free-electron laser}.
\newblock {\em Nature Photonics}, 6(10):693--698, August 2012.

\bibitem{Emma2010}
P.~Emma, R.~Akre, J.~Arthur, R.~Bionta, C.~Bostedt, J.~Bozek, A.~Brachmann,
  P.~Bucksbaum, R.~Coffee, F.-J. Decker, Y.~Ding, D.~Dowell, S.~Edstrom,
  A.~Fisher, J.~Frisch, S.~Gilevich, J.~Hastings, G.~Hays, Ph. Hering,
  Z.~Huang, R.~Iverson, H.~Loos, M.~Messerschmidt, A.~Miahnahri, S.~Moeller,
  H.-D. Nuhn, G.~Pile, D.~Ratner, J.~Rzepiela, D.~Schultz, T.~Smith, P.~Stefan,
  H.~Tompkins, J.~Turner, J.~Welch, W.~White, J.~Wu, G.~Yocky, and J.~Galayda.
\newblock {First lasing and operation of an \aa ngstrom-wavelength
  free-electron laser}.
\newblock {\em Nature Photonics}, 4(9):641--647, August 2010.

\bibitem{Pile2011}
David Pile.
\newblock {X-rays: First light from SACLA}.
\newblock {\em Nature Photonics}, 5(8):456--457, July 2011.

\bibitem{Oppelt2007}
A~Oppelt, A~Adelmann, A~Anghel, R~J Bakker, M~Dehler, R~Ganter, C~Gough,
  S~Ivkovic, F~Jenni, C~Kraus, S~C Leemann, F~Le Pimpec, K~Li, P~Ming,
  B~Oswald, M~Paraliev, M~Pedrozzi, J~Raguin, L~Rivkin, T~Schietinger,
  V~Schlott, L~Schulz, A~Streun, F~Stulle, D~Vermeulen, F~Wei, A~F Wrulich, and
  Villigen Psi.
\newblock {TOWARDS A LOW EMITTANCE X-RAY FEL AT PSI}.
\newblock {\em Proc. of FEL Novosibirsk, Russia}, page 224, 2007.

\bibitem{Schwarz2004}
A.S.Schwarz and {European XFEL Group}.
\newblock {THE EUROPEAN X-RAY FREE ELECTRON LASER PROJECT AT DESY}.
\newblock In {\em Proceedings of the 2004 FEL Conference}, page~85, Trieste,
  Italy, 2004.

\bibitem{Schropp2013}
Andreas Schropp, Robert Hoppe, Vivienne Meier, Jens Patommel, Frank Seiboth,
  Hae~Ja Lee, Bob Nagler, Eric~C. Galtier, Brice Arnold, Ulf Zastrau, Jerome~B.
  Hastings, Daniel Nilsson, Fredrik Uhl\'{e}n, Ulrich Vogt, Hans~M. Hertz, and
  Christian~G. Schroer.
\newblock {Full spatial characterization of a nanofocused x-ray free-electron
  laser beam by ptychographic imaging}.
\newblock {\em Scientific Reports}, 3, April 2013.

\bibitem{Tong2005}
X~M Tong and C~D Lin.
\newblock {Empirical formula for static field ionization rates of atoms and
  molecules by lasers in the barrier-suppression regime}.
\newblock {\em Journal of Physics B: Atomic, Molecular and Optical Physics},
  38(15):2593--2600, August 2005.

\bibitem{Tabak1994}
Max Tabak, James Hammer, Michael~E. Glinsky, William~L. Kruer, Scott~C. Wilks,
  John Woodworth, E.~Michael Campbell, Michael~D. Perry, and Rodney~J. Mason.
\newblock {Ignition and high gain with ultrapowerful lasers}.
\newblock {\em Physics of Plasmas}, 1(5):1626, 1994.

\bibitem{Lindl1995}
John Lindl.
\newblock {Development of the indirect-drive approach to inertial confinement
  fusion and the target physics basis for ignition and gain}.
\newblock {\em Physics of Plasmas}, 2(11):3933, 1995.

\bibitem{Kodama2001}
R~Kodama, P~a Norreys, K~Mima, a~E Dangor, R~G Evans, H~Fujita, Y~Kitagawa,
  K~Krushelnick, T~Miyakoshi, N~Miyanaga, T~Norimatsu, S~J Rose, T~Shozaki,
  K~Shigemori, A~Sunahara, M~Tampo, K~a Tanaka, Y~Toyama, T~Yamanaka, and
  M~Zepf.
\newblock {Fast heating of ultrahigh-density plasma as a step towards laser
  fusion ignition.}
\newblock {\em Nature}, 412(6849):798, August 2001.

\bibitem{MosesFusionNIF}
Edward~I. Moses, Christopher~J. Keane, Rokaya Al-Ayat, Bruce~A. Remington, and
  Gilbert~W. Collins.
\newblock {Inertial confinement fusion and high energy density science
  experiments at the National Ignition Facility}.
\newblock {\em Purazuma, Kaku Yugo Gakkai-Shi}, 87(5):295, 2011.

\bibitem{Pegoranov-PhotonBubbles}
F.~Pegoraro and S.~V. Bulanov.
\newblock {Photon Bubbles and Ion Acceleration in a Plasma Dominated by the
  Radiation Pressure of an Electromagnetic Pulse}.
\newblock {\em Physical Review Letters}, 99(6):65002, August 2007.

\bibitem{Henig2009a}
A.~Henig, S.~Steinke, M.~Schn\"{u}rer, T.~Sokollik, R.~H\"{o}rlein, D.~Kiefer,
  D.~Jung, J.~Schreiber, B.~M. Hegelich, X.~Q. Yan, J.~Meyer-ter Vehn,
  T.~Tajima, P.~V. Nickles, W.~Sandner, D.~Habs, and Others.
\newblock {Radiation-Pressure Acceleration of Ion Beams Driven by Circularly
  Polarized Laser Pulses}.
\newblock {\em Phys. Rev. Lett.}, 103(24):245003, December 2009.

\bibitem{Palmer2012}
C.~A.~J. Palmer, J.~Schreiber, S.~R. Nagel, N.~P. Dover, C.~Bellei, F.~N. Beg,
  S.~Bott, R.~J. Clarke, A.~E. Dangor, S.~M. Hassan, P.~Hilz, D.~Jung,
  S.~Kneip, S.~P.~D. Mangles, K.~L. Lancaster, A.~Rehman, A.~P.~L. Robinson,
  C.~Spindloe, J.~Szerypo, M.~Tatarakis, M.~Yeung, M.~Zepf, and Z.~Najmudin.
\newblock {Rayleigh-Taylor Instability of an Ultrathin Foil Accelerated by the
  Radiation Pressure of an Intense Laser}.
\newblock {\em Physical Review Letters}, 108(22):225002, May 2012.

\bibitem{Steinke2013}
S.~Steinke, P.~Hilz, M.~Schn\"{u}rer, G.~Priebe, J.~Br\"{a}nzel, F.~Abicht,
  D.~Kiefer, C.~Kreuzer, T.~Ostermayr, J.~Schreiber, A.~A. Andreev, T.~P. Yu,
  A.~Pukhov, and W.~Sandner.
\newblock {Stable laser-ion acceleration in the light sail regime}.
\newblock {\em Physical Review Special Topics - Accelerators and Beams},
  16(1):011303, January 2013.

\bibitem{Wilks2001}
S.~C. Wilks, A.~B. Langdon, T.~E. Cowan, M.~Roth, M.~Singh, S.~Hatchett, M.~H.
  Key, D.~Pennington, A.~MacKinnon, and R.~A. Snavely.
\newblock {Energetic proton generation in ultra-intense laser–solid
  interactions}.
\newblock {\em Physics of Plasmas}, 8(2):542, February 2001.

\bibitem{Zeil2010}
K~Zeil, S~D Kraft, S~Bock, M~Bussmann, T~E Cowan, T~Kluge, J~Metzkes,
  T~Richter, R~Sauerbrey, and U~Schramm.
\newblock {The scaling of proton energies in ultrashort pulse laser plasma
  acceleration}.
\newblock {\em New Journal of Physics}, 12(4):045015, April 2010.

\bibitem{Gaillard2011}
S~A Gaillard, T~Kluge, K~A Flippo, M~Bussmann, B~Gall, T~Lockard, M~Geissel,
  D~T Offermann, M~Schollmeier, Y~Sentoku, and T~E Cowan.
\newblock {Increased laser-accelerated proton energies via direct
  laser-light-pressure acceleration of electrons in microcone targets}.
\newblock {\em Physics of Plasmas}, 18(5):056710, 2011.

\bibitem{Perego2012}
C~Perego, D~Batani, a~Zani, and M~Passoni.
\newblock {Target normal sheath acceleration analytical modeling, comparative
  study and developments.}
\newblock {\em The Review of scientific instruments}, 83(2):02B502, February
  2012.

\bibitem{Kim}
I~Jong Kim, Ki~Hong Pae, Chul~Min Kim, Hyung~Taek Kim, Jae~Hee Sung, and
  Seong~Ku Lee.
\newblock {Towards radiation pressure acceleration of protons using linearly
  polarized ultrashort petawatt laser pulses}.
\newblock {\em arXiv preprint}, 2013.

\bibitem{Forslund1971}
D.~W. Forslund and J.~P. Freidberg.
\newblock {Theory of Laminar Collisionless Shocks}.
\newblock {\em Phys. Rev. Lett.}, 27(18):1189, 1971.

\bibitem{Silva2004}
Lu\'{\i}s O. Lu$\backslash$'$\backslash$is~O Silva, Michael Marti, Jonathan~R.
  Davies, Ricardo~a. Fonseca, Chuang Ren, Frank~S Tsung, and Warren~B Mori.
\newblock {Proton Shock Acceleration in Laser-Plasma Interactions}.
\newblock {\em Phys. Rev. Lett.}, 92(1):15002, January 2004.

\bibitem{DHumieres2005}
E~D'Humieres, Others, Emmanuel D’Humières, Erik Lefebvre, Laurent
  Gremillet, and Victor Malka.
\newblock {Proton acceleration mechanisms in high-intensity laser interaction
  with thin foils}.
\newblock {\em Physics of Plasmas}, 12(6):062704, 2005.

\bibitem{Fiuza2012}
F.~Fiuza, a.~Stockem, E.~Boella, R.~a. Fonseca, L.~O. Silva, D.~Haberberger,
  S.~Tochitsky, C.~Gong, W.~B. Mori, and C.~Joshi.
\newblock {Laser-Driven Shock Acceleration of Monoenergetic Ion Beams}.
\newblock {\em Physical Review Letters}, 109(21):215001, November 2012.

\bibitem{Sentoku2007}
Y.~Sentoku, a.~J. Kemp, R.~Presura, M.~S. Bakeman, and T.~E. Cowan.
\newblock {Isochoric heating in heterogeneous solid targets with ultrashort
  laser pulses}.
\newblock {\em Physics of Plasmas}, 14(12):122701, 2007.

\bibitem{Akli2008}
K.~Akli, S.~Hansen, a.~Kemp, R.~Freeman, F.~Beg, D.~Clark, S.~Chen, D.~Hey,
  S.~Hatchett, K.~Highbarger, E.~Giraldez, J.~Green, G.~Gregori, K.~Lancaster,
  T.~Ma, a.~MacKinnon, P.~Norreys, N.~Patel, J.~Pasley, C.~Shearer,
  R.~Stephens, C.~Stoeckl, M.~Storm, W.~Theobald, L.~{Van Woerkom}, R.~Weber,
  and M.~Key.
\newblock {Laser Heating of Solid Matter by Light-Pressure-Driven Shocks at
  Ultrarelativistic Intensities}.
\newblock {\em Physical Review Letters}, 100(16):165002, April 2008.

\bibitem{LGHuang}
{L. G. Huang}, T.~Kluge, M.~Bussmann, and T.~E. Cowan~et al.
\newblock {Bulk ion heating in an optically thin buried layer target irradiated
  by an ultra-short intense laser pulse}.
\newblock {\em (in preparation)}.

\bibitem{HHGSW}
Daniel an~der Br\"{u}gge, Naveen Kumar, Alexander Pukhov, and Christian
  R\"{o}del.
\newblock {Influence of Surface Waves on Plasma High-Order Harmonic
  Generation}.
\newblock {\em Physical Review Letters}, 108(12):125002, March 2012.

\bibitem{USDepartmentofCommerce}
J.~H. Hubbell and S.~M. Seltzer.
\newblock {NIST: X-Ray Mass Attenuation Coefficients (Standard Reference
  Database 126)}.

\bibitem{McNulty1992}
I~McNulty, J~Kirz, C~Jacobsen, E~H Anderson, M~R Howells, and D~P Kern.
\newblock {High-Resolution Imaging by Fourier Transform X-ray Holography.}
\newblock {\em Science (New York, N.Y.)}, 256(5059):1009--12, May 1992.

\bibitem{Lindaas1996}
Steve Lindaas, Malcolm Howells, Chris Jacobsen, and Alex Kalinovsky.
\newblock {X-ray holographic microscopy by means of photoresist recording and
  atomic-force microscope readout}.
\newblock {\em Journal of the Optical Society of America A}, 13(9):1788,
  September 1996.

\bibitem{Mancuso2010}
A.~P. Mancuso, Th. Gorniak, F.~Staier, O.~M. Yefanov, R.~Barth, C.~Christophis,
  B.~Reime, J.~Gulden, A.~Singer, M.~E. Pettit, Th. Nisius, Th. Wilhein,
  C.~Gutt, G.~Gr\"{u}bel, N.~Guerassimova, R.~Treusch, J~Feldhaus, S.~Eisebitt,
  E.~Weckert, M.~Grunze, A.~Rosenhahn, and I.~A. Vartanyants.
\newblock {Coherent imaging of biological samples with femtosecond pulses at
  the free-electron laser FLASH}.
\newblock {\em New Journal of Physics}, 12(3):035003, March 2010.

\bibitem{Miao1999}
Jianwei Miao, Pambos Charalambous, Janos Kirz, and David Sayre.
\newblock {Extending the methodology of X-ray crystallography to allow imaging
  of micrometre-sized non-crystalline specimens}.
\newblock {\em Nature}, 400(6742):342, July 1999.

\bibitem{Fienup1982}
J.~R. Fienup.
\newblock {Phase retrieval algorithms: a comparison}.
\newblock {\em Applied Optics}, 21(15):2758, August 1982.

\bibitem{RCXDI}
Changyong Song, Raymond Bergstrom, Damien Ramunno-Johnson, Huaidong Jiang,
  David Paterson, Martin de~Jonge, Ian McNulty, Jooyoung Lee, Kang Wang, and
  Jianwei Miao.
\newblock {Nanoscale Imaging of Buried Structures with Elemental Specificity
  Using Resonant X-Ray Diffraction Microscopy}.
\newblock {\em Physical Review Letters}, 100(2):025504, January 2008.

\bibitem{Son2011}
Sang-Kil Son, Henry~N. Chapman, and Robin Santra.
\newblock {Multiwavelength Anomalous Diffraction at High X-Ray Intensity}.
\newblock {\em Physical Review Letters}, 107(21):218102, November 2011.

\bibitem{picls}
Y~Sentoku and A~J Kemp.
\newblock {Numerical methods for particle simulations at extreme densities and
  temperatures: Weighted particles, relativistic collisions and reduced
  currents}.
\newblock {\em Journal of Computational Physics}, 227:6846, 2008.

\bibitem{Kemp2008}
A.~Kemp, Y.~Sentoku, and M.~Tabak.
\newblock {Hot-Electron Energy Coupling in Ultraintense Laser-Matter
  Interaction}.
\newblock {\em Physical Review Letters}, 101(7):075004, August 2008.

\bibitem{Chrisman2008}
B.~Chrisman, Y.~Sentoku, and a.~J. Kemp.
\newblock {Intensity scaling of hot electron energy coupling in cone-guided
  fast ignition}.
\newblock {\em Physics of Plasmas}, 15(5):056309, 2008.

\bibitem{Mishra2009}
R.~Mishra, Y.~Sentoku, and a.~J. Kemp.
\newblock {Hot electron generation forming a steep interface in superintense
  laser-matter interaction}.
\newblock {\em Physics of Plasmas}, 16(11):112704, 2009.

\bibitem{Gibbon1992}
Paul Gibbon.
\newblock {Collisionless Absorption in Sharp-Edged Plasmas}.
\newblock {\em Phys. Rev. Lett.}, 68(10):1535, 1992.

\bibitem{Mulser2008}
P.~Mulser, D.~Bauer, and H.~Ruhl.
\newblock {Collisionless Laser-Energy Conversion by Anharmonic Resonance}.
\newblock {\em Physical Review Letters}, 101(22):225002, November 2008.

\bibitem{Kluge2011a}
T~Kluge, T~E Cowan, A~Debus, U~Schramm, K~Zeil, M~Bussmann, A~Others,
  Helmholtzzentrum Dresden-rossendorf V, and Bautzner Landstra\ss~e.
\newblock {Electron Temperature Scaling in Laser Interaction with Solid Foils}.
\newblock {\em Phys. Rev. Lett.}, 107(20):205003, 2011.

\bibitem{Rodel2012}
C.~R\"{o}del, D.~an~der Br\"{u}gge, J.~Bierbach, M.~Yeung, T.~Hahn, B.~Dromey,
  S.~Herzer, S.~Fuchs, A.~Galestian Pour, E.~Eckner, M.~Behmke, M.~Cerchez,
  O.~J\"{a}ckel, D.~Hemmers, T.~Toncian, M.~C. Kaluza, A.~Belyanin,
  G.~Pretzler, O.~Willi, A.~Pukhov, M.~Zepf, and G.~G. Paulus.
\newblock {Harmonic Generation from Relativistic Plasma Surfaces in Ultrasteep
  Plasma Density Gradients}.
\newblock {\em Physical Review Letters}, 109(12):125002, September 2012.

\bibitem{Sentoku2000}
Yasuhiko Sentoku, Kunioki Mima, Shin-ichi Kojima, and Hartmut Ruhl.
\newblock {Magnetic instability by the relativistic laser pulses in overdense
  plasmas}.
\newblock {\em Physics of Plasmas}, 7(2):689, February 2000.

\bibitem{Macchi2001}
A.~Macchi, F.~Cornolti, F.~Pegoraro, T.~Liseikina, H.~Ruhl, and V.~Vshivkov.
\newblock {Surface Oscillations in Overdense Plasmas Irradiated by Ultrashort
  Laser Pulses}.
\newblock {\em Physical Review Letters}, 87(20):205004, October 2001.

\bibitem{Macchi2002}
Andrea Macchi, Fulvio Cornolti, and Francesco Pegoraro.
\newblock {Two-surface wave decay}.
\newblock {\em Physics of Plasmas}, 9(5):1704, May 2002.

\bibitem{Kumar2007}
Naveen Kumar and V.~K. Tripathi.
\newblock {Parametric excitation of surface plasma waves in an overdense plasma
  irradiated by an ultrashort laser pulse}.
\newblock {\em Physics of Plasmas}, 14(10):103108, October 2007.

\bibitem{Sentokuresistive}
Y.~Sentoku, E.~D’Humi\`{e}res, L.~Romagnani, P.~Audebert, and J.~Fuchs.
\newblock {Dynamic Control over Mega-Ampere Electron Currents in Metals Using
  Ionization-Driven Resistive Magnetic Fields}.
\newblock {\em Physical Review Letters}, 107(13):135005, September 2011.

\bibitem{HainesInstability}
M.~G. Haines.
\newblock {Thermal Instability and Magnetic Field Generated by Large Heat Flow
  in a Plasma, Especially under Laser-Fusion Conditions}.
\newblock {\em Phys. Rev. Lett.}, 47(13):917, 1981.

\bibitem{Fuchs2003}
J.~Fuchs, T.~Cowan, P.~Audebert, H.~Ruhl, L.~Gremillet, a.~Kemp, M.~Allen,
  a.~Blazevic, J.-C. Gauthier, M.~Geissel, M.~Hegelich, S.~Karsch, P.~Parks,
  M.~Roth, Y.~Sentoku, R.~Stephens, and E.~Campbell.
\newblock {Spatial Uniformity of Laser-Accelerated Ultrahigh-Current MeV
  Electron Propagation in Metals and Insulators}.
\newblock {\em Physical Review Letters}, 91(25):255002, December 2003.

\bibitem{Metzkes}
J.~Metzkes~et al.
\newblock {Filamentation in thin targets}.
\newblock {\em (in preparation)}.

\bibitem{Grubel2007}
G.~Gr\"{u}bel, G.B. Stephenson, C.~Gutt, H.~Sinn, and Th. Tschentscher.
\newblock {XPCS at the European X-ray free electron laser facility}.
\newblock {\em Nuclear Instruments and Methods in Physics Research Section B:
  Beam Interactions with Materials and Atoms}, 262(2):357, September 2007.

\bibitem{Gutt2008}
C.~Gutt, L.-M. Stadler, A.~Duri, T.~Autenrieth, O.~Leupold, Y.~Chushkin, and
  G.~Gr\"{u}bel.
\newblock {Measuring temporal speckle correlations at ultrafast x-ray sources}.
\newblock {\em Optics Express}, 17(1):55, December 2008.

\bibitem{Tschentscher2011}
Th. Tschentscher.
\newblock {Layout of the X-Ray Systems at the European XFEL}.
\newblock Technical report, DESY, Hamburg, 2011.

\bibitem{Neumayer2010}
P.~Neumayer, B.~Aurand, M.~Basko, B.~Ecker, P.~Gibbon, D.~C. Hochhaus,
  A.~Karmakar, E.~Kazakov, T.~Kühl, C.~Labaune, O.~Rosmej, An. Tauschwitz,
  B.~Zielbauer, and D.~Zimmer.
\newblock {The role of hot electron refluxing in laser-generated K-alpha
  sources}.
\newblock {\em Physics of Plasmas}, 17(10):103103, October 2010.

\bibitem{HIBEFConsortium}
{HIBEF Consortium}.
\newblock http://www.hzdr.de/hgfbeamline.

\end{thebibliography}
\end{document}